\documentclass[aps,prl,twocolumn,nofootinbib,a4paper,amsfonts,amssymb,amsmath,floatfix,superscriptaddress]{revtex4-2}
\pdfoutput=1
\usepackage[utf8]{inputenc}
\usepackage[english]{babel}
\usepackage[colorlinks=true,allcolors=magenta,breaklinks=true]{hyperref}
\usepackage{graphicx}
\usepackage{mathtools}
\usepackage{dsfont}
\usepackage{tikz}
\usetikzlibrary{calc,shapes,patterns,decorations.pathreplacing,calligraphy,svg.path}
\usepackage{soul}
\usepackage{microtype}
\usepackage{subcaption}
\usepackage{comment}
\usepackage{microtype}

\newcommand{\ket}[1]{|#1\rangle}                        
\newcommand{\bra}[1]{\langle #1|}                       
\newcommand{\braket}[2]{\langle #1 | #2 \rangle}        

\usepackage[normalem]{ulem}
\begin{document}

\title{Trading causal order for locality}

\author{Ravi Kunjwal}
	\affiliation{Centre for Quantum Information and Communication (QuIC), Ecole polytechnique de Bruxelles, CP 165, Universit\'e libre de Bruxelles, 1050 Brussels, Belgium}

\author{\"Amin Baumeler}
	\affiliation{Facolt\`a di scienze informatiche, Universit\`a della Svizzera italiana, 6900 Lugano, Switzerland}
	\affiliation{Facolt\`a indipendente di Gandria, 6978 Gandria, Switzerland}

\date{\today}                                           

\begin{abstract}
	\noindent
	Quantum theory admits ensembles of quantum nonlocality without entanglement (QNLWE).
	These ensembles consist of seemingly classical states (they are perfectly distinguishable and non-entangled) that cannot be perfectly discriminated with local operations and classical communication (LOCC).
	Here, we analyze QNLWE from a causal perspective, and show how to perfectly discriminate some of these ensembles using local operations and classical communication {\em without definite causal order.} 
	Specifically, three parties with access to an instance of indefinite causal order---the AF/BW process---can perfectly discriminate the states in a QNLWE ensemble---the SHIFT ensemble---with {\em local\/} operations.
	Hence, this type of quantum nonlocality disappears at the expense of definite causal order while retaining classical communication. Our results thereby leverage the fact that LOCC is a conjunction of \textit{three} constraints: local operations, classical communication, and definite causal order.
	Moreover, we show how multipartite generalizations of the AF/BW process are transformed into multiqubit ensembles that exhibit QNLWE.
	Such ensembles are of independent interest for cryptographic protocols and for the study of separable quantum operations unachievable with LOCC.
\end{abstract}

\maketitle

\textit{Introduction.---}The famously counter-intuitive nature of quantum theory owes much to the phenomenon of entanglement which forces {\em ``its entire departure from classical lines of thought''\/} \cite{schrodinger1935c}.
Perhaps the deepest consequence of entanglement is its role in revealing the tension between quantum theory and locality which is central to Bell's theorem~\cite{bell1964}.
This tension, however, does not stop at entanglement and Bell's theorem:
It persists in a different form even without entanglement, as captured by the phenomenon of quantum nonlocality without entanglement~(QNLWE)~\cite{bennett1999}.

At the heart of quantum nonlocality---with or without entanglement---is the interplay of causation and correlation \cite{wood2015,WSS20}.\footnote{Note that the notion of locality at play in Bell's theorem is local causality \cite{wiseman2014} while in QNLWE, the notion of locality at play is the locality of operations \cite{bennett1999}.}
To demonstrate QNLWE, Bennett~{\it et al.}~\cite{bennett1999} present 
{\em locally\/} imperfectly discriminable ensembles of mutually orthogonal product quantum states,~{\it e.g.,\/}~the SHIFT ensemble
\begin{align}
	\begin{split}
		\{&\ket{000},\ket{111},\ket{{+}01},\ket{{-}01},\\&\ket{1{+}0},\ket{1{-}0},\ket{01{+}},\ket{01{-}}\}
		\,.
	\end{split}
	\label{eq:shift}
\end{align}
Although states in such an ensemble can be prepared locally, parties sharing an unknown state from the ensemble cannot perfectly identify the state with local operations and classical communication~(LOCC).
The classical communication in LOCC is implicitly assumed to respect a~{\em definite causal order\/} (`causal order' for short):\footnote{Unless specified otherwise, `causal order' will always mean `definite causal order' in this paper, in keeping with Ref.~\cite{oreshkov2012}.}
In each round, the direction of communication is determined from all past data.
A~necessary consequence of this constraint is that at least one party must initiate the communication.
By contrast, in the case of Bell nonlocality, all communication is excluded by the requirement of spacelike separation.
The background assumption of definite causal order, however, is common to both types of nonlocality.

What if we drop the assumption of a definite causal order and regard it as a physical quantity sensitive to quantum indefiniteness~\cite{hardy2005}?
This possibility has attracted much interest in recent years, {\it e.g.,\/}~as in the {\em quantum switch\/}~\cite{chiribella2013} achievable through indefinite wires connecting quantum gates~\cite{colnaghi2012} or through indefinite spacetime geometries formed from matter in a superposition of locations~\cite{zych2017}.
Oreshkov, Costa, and Brukner~\cite{oreshkov2012} show that if, without further assumptions on causal connections, one insists that parties locally cannot detect any deviation from standard quantum theory, then indefinite causal order arises naturally:
Their process-matrix framework encompasses the quantum switch~\cite{og16,araujo2015}, and also exhibits {\em noncausal correlations,\/}~{\it i.e.,\/}~correlations unattainable under a global causal order among the parties (see also Refs.~\cite{baumeler2014ieee,branciard2015,abbott2016}).
Moreover, they show that the exotic causal possibilities that arise between two parties disappear in the classical limit.

For three parties or more, however, {\em logically consistent classical processes\/} that create noncausal correlations exist~\cite{baumeler2014} (the interested reader may consult the Appendix for more details).
For example, the deterministic Araújo-Feix/Baumeler-Wolf (AF/BW) process~\cite{af,baumeler2016} exchanges bits among three parties, Alice, Bob, and Charlie, in the following way.
Each party receives a~bit
\begin{align}
	a:=(y\oplus 1)z,\quad
	b:=(z\oplus 1)x,\quad
	c:=(x\oplus 1)y
	\label{eq:afbw}
\end{align}
from the process and {\em thereafter\/} provides a bit of their choice~$x,y,z$ to the process.
This resource allows every pair of parties to communicate to the third ({\it e.g.,\/}~Alice receives~$a$ which non-trivially depends on~$y,z$ of Bob and Charlie) in a single round:
{\em Each party acts in the causal future of the other two.\/}

The possibility of indefinite causal order---in particular, the AF/BW process---raises the following natural question:
What happens to the tension between quantum theory and locality once the assumption of definite causal order is dropped?

\textit{Results.---}In this Letter, we show how one can trade causal order for the locality of operations in perfectly discriminating QNLWE ensembles.
The tension between quantum theory and locality suggested by QNLWE thus disappears in the absence of a definite causal order.
Specifically, {\em local quantum\/} operations assisted with {\em classical processes\/} can allow the parties
to {\em perfectly discriminate ensembles of quantum nonlocality without entanglement:\/}
Three parties communicating through the classical AF/BW process (Eq.~\eqref{eq:afbw}) can discriminate the SHIFT ensemble~(Eq.~\eqref{eq:shift}).
In fact, this process allows the parties to {\em measure\/} quantum systems in the SHIFT basis.
Conversely, we show that such a measurement implements the classical channel underlying the AF/BW process.
We use the insights from these protocols to show how any Boolean~$n$-party classical process without global past can be turned into an~$n$-qubit ensemble of states that exhibits quantum nonlocality without entanglement.
These results establish an {\em operational link\/} between QNLWE and classical processes without causal order.\footnote{See also Ref.~\cite{baumeler2022} for a suggested link between such processes and Bell nonlocality and Ref.~\cite{vanderlugt2022} for a tension between the assumptions of definite causal order and parameter independence.}

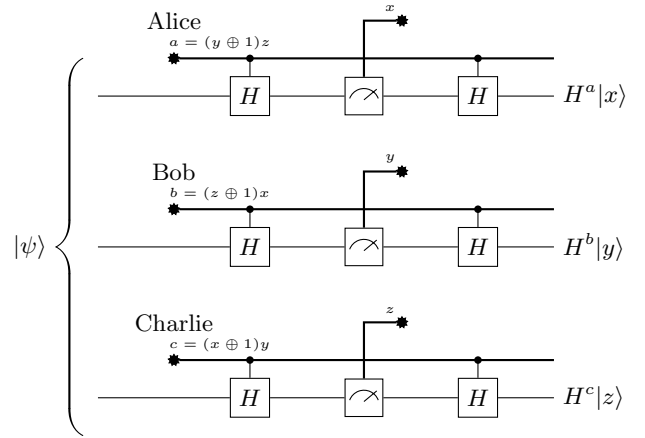
\begin{figure}
	\centering
	\begin{tikzpicture}
		\node (AL) at (0,4) {Alice};
		\node (BL) at (0,2) {Bob};
		\node (CL) at (0,0) {Charlie};
		\node[draw,star,star points=9,inner sep=1,fill=black] at ($ (AL) + (0,-.5) $) (a) {};
		\node[draw,star,star points=9,inner sep=1,fill=black] at ($ (BL) + (0,-.5) $) (b) {};
		\node[draw,star,star points=9,inner sep=1,fill=black] at ($ (CL) + (0,-.5) $) (c) {};
		\node[draw,star,star points=9,inner sep=1,fill=black] at ($ (AL) + (3,0) $) (x) {};
		\node[draw,star,star points=9,inner sep=1,fill=black] at ($ (BL) + (3,0) $) (y) {};
		\node[draw,star,star points=9,inner sep=1,fill=black] at ($ (CL) + (3,0) $) (z) {};
		\node[draw,regular polygon,regular polygon sides=4,inner sep=1,minimum size=20] at ($ (AL) + (1,-1) $) (hA) {$H$};
		\node[draw,regular polygon,regular polygon sides=4,inner sep=1,minimum size=20] at ($ (BL) + (1,-1) $) (hB) {$H$};
		\node[draw,regular polygon,regular polygon sides=4,inner sep=1,minimum size=20] at ($ (CL) + (1,-1) $) (hC) {$H$};
		\draw (hA) -- ++(0,.5) node[circle,inner sep=1,fill=black] {};
		\draw (hB) -- ++(0,.5) node[circle,inner sep=1,fill=black] {};
		\draw (hC) -- ++(0,.5) node[circle,inner sep=1,fill=black] {};
		\node[draw,regular polygon,regular polygon sides=4,inner sep=1,minimum size=20] at ($ (AL) + (2.5,-1) $) (mA) {};
		\node[draw,regular polygon,regular polygon sides=4,inner sep=1,minimum size=20] at ($ (BL) + (2.5,-1) $) (mB) {};
		\node[draw,regular polygon,regular polygon sides=4,inner sep=1,minimum size=20] at ($ (CL) + (2.5,-1) $) (mC) {};
		\draw ($ (mA) - (0,0.1) $) ++(20:0.2) arc (20:160:0.2) ($ (mA) - (0,0.05) $) -- ++(40:0.22);
		\draw ($ (mB) - (0,0.1) $) ++(20:0.2) arc (20:160:0.2) ($ (mB) - (0,0.05) $) -- ++(40:0.22);
		\draw ($ (mC) - (0,0.1) $) ++(20:0.2) arc (20:160:0.2) ($ (mC) - (0,0.05) $) -- ++(40:0.22);
		\node[draw,regular polygon,regular polygon sides=4,inner sep=1,minimum size=20] at ($ (AL) + (4,-1) $) (hA2) {$H$};
		\node[draw,regular polygon,regular polygon sides=4,inner sep=1,minimum size=20] at ($ (BL) + (4,-1) $) (hB2) {$H$};
		\node[draw,regular polygon,regular polygon sides=4,inner sep=1,minimum size=20] at ($ (CL) + (4,-1) $) (hC2) {$H$};
		\draw (hA2) -- ++(0,.5) node[circle,inner sep=1,fill=black] {};
		\draw (hB2) -- ++(0,.5) node[circle,inner sep=1,fill=black] {};
		\draw (hC2) -- ++(0,.5) node[circle,inner sep=1,fill=black] {};
		\draw[thick] (a) -- ++(5,0) node[pos=.108,above] {\tiny $a=(y\oplus 1)z$};
		\draw[thick] (b) -- ++(5,0) node[pos=.108,above] {\tiny $b=(z\oplus 1)x$};
		\draw[thick] (c) -- ++(5,0) node[pos=.108,above] {\tiny $c=(x\oplus 1)y$};
		\draw[thick] (mA) |- (x) node[pos=.9,above] {\tiny $x$};
		\draw[thick] (mB) |- (y) node[pos=.9,above] {\tiny $y$};
		\draw[thick] (mC) |- (z) node[pos=.9,above] {\tiny $z$};
		\draw ($ (hA) - (2,0) $) -- (hA) (hA) -- (mA) (mA) -- (hA2)  (hA2) -- ++(1,0) node[right] {$H^a\ket x$};
		\draw ($ (hB) - (2,0) $) -- (hB) (hB) -- (mB) (mB) -- (hB2)  (hB2) -- ++(1,0) node[right] {$H^b\ket y$};
		\draw ($ (hC) - (2,0) $) -- (hC) (hC) -- (mC) (mC) -- (hC2)  (hC2) -- ++(1,0) node[right] {$H^c\ket z$};
		\draw[decorate,decoration={calligraphic brace,amplitude=10pt},thick] ($ (hC) + (-2.2,-.5) $) -- ($ (hA) + (-2.2,.5) $) node[midway,left=10pt] {$\ket\psi$};
	\end{tikzpicture}
	\caption{Schematic of protocol to implement the SHIFT-basis measurement on an arbitrary quantum state~$\ket\psi$ with local operations and classical communication without causal order.
		Thick wires represent classical bits, normal wires qubits, and \protect\tikz{\protect\node[draw,star,star points=9,inner sep=1,fill=black] at (0,0) {};} represents the interface to the AF/BW process.
	}
	\label{fig:measurementprotocol}
\end{figure}

\textit{SHIFT-basis measurement from AF/BW process.---}The parties Alice, Bob, and Charlie hold a quantum system in the three-qubit state~$\ket\psi$.
The following protocol implements a measurement of~$\ket\psi$ in the SHIFT basis~(see~Fig.~\ref{fig:measurementprotocol}).
First, each party receives a~classical bit~$a,b,c$ from the process.
Then, each party applies a Hadamard transformation on their share of~$\ket\psi$ if the received bit is~$1$, {\it i.e.,\/}~the parties apply~\mbox{$H^{(a,b,c)}:=H^a\otimes H^b\otimes H^c$}.
Now, they measure the~quantum system in the computational basis, obtain~the post-measurement state~$\ket{xyz}$, and forward~$x,y,z$ to the AF/BW process.
Finally, the parties apply~$H^{(a,b,c)}$ to the post-measurement state.
By this, the final state of the quantum system is
\begin{align}
	\sum_{x,y,z}
	\left|\bra{xyz}H^{(a,b,c)}\ket\psi\right|^2
	H^{(a,b,c)}
	\ket{xyz}
	\bra{xyz}
	H^{(a,b,c)}
	\,.
	\label{eq:final}
\end{align}
Note that the AF/BW process determines the values of~$a,b,c$ as a function of~$x,y,z$.

First, we show that if~$\ket\psi\in\text{SHIFT}$, then this protocol returns the state~$\ket\psi$ with certainty.
If~$\ket\psi=\ket{000}$, then the probability
\begin{align}
	|\bra{xyz}H^{((y\oplus 1)z,(z\oplus 1)x,(x\oplus 1)y)}\ket{000}|^2
\end{align}
is one for~$x=y=z=0$, and zero otherwise:
The final state is~$\ket{000}$.
Instead, if~$\ket\psi=\ket{01+}$, then the only contribution arises for~$x=z=0$ and~$y=1$~$(|\bra{010}H^{(0,0,1)}\ket{01+}|^2 = 1)$,
and the final state is~$H^{(0,0,1)}\ket{010}=\ket{01{+}}$.
By symmetry, the same follows for all SHIFT-ensemble states.
In other words, for each SHIFT state there exists a {\em unique and distinct\/} triple~$x,y,z$ that contributes to the sum;
namely,~$x,y,z$ encode the qubits of the SHIFT state~($0$~if the qubit is in the state~$\ket 0$ or~$\ket +$, and~$1$ otherwise).
By linearity, this analysis extends to {\em any\/} quantum state~$\ket\psi$:
Measuring an arbitrary state~$\ket\psi=\sum_{\ket k\in\text{SHIFT}}\alpha_{\ket k} \ket k$ in the SHIFT basis yields~$\sum_{\ket k\in\text{SHIFT}}|\alpha_{\ket k}|^2\ket k\bra k$, which is identical to the returned state of the protocol
\begin{align}
	&|\alpha_{\ket{000}}|^2 \ket{000}\bra{000}\\
	&\qquad+
	|\alpha_{\ket{{+}01}}|^2 H^{(1,0,0)}\ket{001}\bra{001}H^{(1,0,0)}\\
	&\qquad+
	|\alpha_{\ket{01{+}}}|^2 H^{(0,0,1)}\ket{010}\bra{010}H^{(0,0,1)}\\
	&\qquad+\cdots
	\,.
\end{align}

Now it is clear that if the parties communicate through the AF/BW process, then they perfectly discriminate the SHIFT ensemble.
The classical data collected in the above protocol uniquely specifies the SHIFT state they were given:
The bits~$a,b,c$ they receive from the process specify the basis, and the bits~$x,y,z$ they receive from the measurement specify the state in the corresponding basis,
{\it e.g.,\/}~$a=0,b=0,c=1,x=0,y=1,z=0$ encode the state~$\ket{01+}$.

\begin{figure}
	\centering
	\begin{tikzpicture}
		\node (AL) at (0,2) {Alice};
		\node (BL) at (0,1) {Bob};
		\node (CL) at (0,0) {Charlie};
		\node[draw,regular polygon,regular polygon sides=3,inner sep=1,minimum size=10,rotate=30] at ($ (AL) + (.5,-.5) $) (cA) {};
		\node[draw,regular polygon,regular polygon sides=3,inner sep=1,minimum size=10,rotate=30] at ($ (BL) + (.5,-.5) $) (cB) {};
		\node[draw,regular polygon,regular polygon sides=3,inner sep=1,minimum size=10,rotate=30] at ($ (CL) + (.5,-.5) $) (cC) {};
		\node[draw,regular polygon,regular polygon sides=4,inner sep=1,minimum size=20] at ($ (AL) + (4.5,-.5) $) (gA) {$f$};
		\node[draw,regular polygon,regular polygon sides=4,inner sep=1,minimum size=20] at ($ (BL) + (4.5,-.5) $) (gB) {$f$};
		\node[draw,regular polygon,regular polygon sides=4,inner sep=1,minimum size=20] at ($ (CL) + (4.5,-.5) $) (gC) {$f$};
		\draw[thick] (cA) -- ++(-1,0) node[left] {$x$};
		\draw[thick] (cB) -- ++(-1,0) node[left] {$y$};
		\draw[thick] (cC) -- ++(-1,0) node[left] {$z$};
		\draw (cA) -- ++(2,0) node[pos=.2,above] {$\ket x$};
		\draw (cB) -- ++(2,0) node[pos=.2,above] {$\ket y$};
		\draw (cC) -- ++(2,0) node[pos=.2,above] {$\ket z$};
		\draw[thick] ($ (cA) + (2,0) $) -- (gA) node[pos=.8,above] {$\ell_A$};
		\draw[thick] ($ (cB) + (2,0) $) -- (gB) node[pos=.8,above] {$\ell_B$};
		\draw[thick] ($ (cC) + (2,0) $) -- (gC) node[pos=.8,above] {$\ell_C$};
		\draw[thick] (gA) -- ++(1,0) node[right] {$a=(y\oplus 1)z$};
		\draw[thick] (gB) -- ++(1,0) node[right] {$b=(z\oplus 1)x$};
		\draw[thick] (gC) -- ++(1,0) node[right] {$c=(x\oplus 1)y$};
		\draw[fill=white] ($ (cA) + (2,0) + (-.5,.2) $) rectangle ($ (cC) + (2,0) + (.5,-.2) $);
		\draw ($ (cB) + (2,0) - (0,.1) $) ++(20:.3) arc (20:160:.3);
		\draw ($ (cB) + (2,0) - (0,.1) $) -- ++(50:.4);
		\node (sL) at ($ (cB) + (2,0) - (0,.3) $) {\tiny SHIFT};
	\end{tikzpicture}
	\caption{Schematic of protocol to realize the AF/BW channel from a SHIFT-basis measurement.}
	\label{fig:afbwprocessprotocol}
\end{figure}
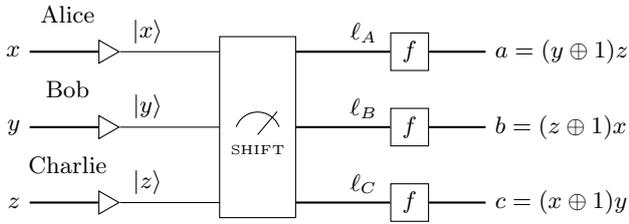

\textit{AF/BW channel from SHIFT-basis measurement.---}
Conversely, suppose three parties, Alice, Bob, and Charlie, have access to a measurement device that measures a three-qubit system in the SHIFT basis and returns to each party the classical description of the post-measurement qubit state.
For instance, if the three-qubit post-measurement state is~$\ket{{+}01}$, then Alice receives the label~$+$, Bob~$0$, and Charlie~$1$.
The following protocol~(see~Fig.~\ref{fig:afbwprocessprotocol}) {\em implements\/} the classical channel underlying the AF/BW process via such a SHIFT-basis measurement, {\it i.e.,\/}~the parties start with three bits~$x,y,z$ of their choice and end up with~$a=(y\oplus 1)z,\allowbreak b=(z\oplus 1)x,c=(x\oplus 1)y$.\footnote{If the variables~$a,b,c$ were in the respective {\em local pasts\/} of the variables~$x,y,z$---as they are in the complementary protocol of Fig.~\ref{fig:measurementprotocol}---this AF/BW {\em channel\/} would correspond to the noncausal AF/BW {\em process.}}

First, each party encodes the respective bit in the computational basis of a qubit, {\it i.e.,\/}~they locally generate a~quantum system in the state~$\ket\psi=\ket{xyz}$.
In the second step, they feed~$\ket\psi$ into the measurement device and record the outcome~$\ell_A,\ell_B,\ell_C\in\{0,1,{+},{-}\}$, where~$\ell_A$ is Alice's outcome and so forth.
Finally, they apply the function~$f:0\mapsto 0,\allowbreak 1\mapsto 0,+\mapsto 1,-\mapsto 1$ to obtain the bits~$a,b,c$.

Suppose the bits~$x,y,z$ are chosen such that~$x=y=z$.
The prepared quantum state~$\ket{xyz}$ is a member of the SHIFT basis.
The measurement device therefore replies the labels~$\ell_A=\ell_B=\ell_C\in\{0,1\}$, and, according to the protocol, the parties set~$a=b=c=0$, which is the correct value.
If the bits are specified as $x=y=0$ and~$z=1$, then the prepared state~$\ket{xyz}$ in not a member of the SHIFT ensemble and the measurement device responds probabilistically:~$|\braket{{+}01}{001}|^2 = |\braket{{-}01}{001}|^2 = 1/2$.
In either case, however, the parties correctly end up with~\mbox{$a=1$},~$b=c=0$.
By symmetry, the parties compute~$a,b,c$ as desired for all inputs~$x,y,z$.

The correspondence between the SHIFT ensemble and the AF/BW process that we have shown above can be understood as a consequence of the following mathematical fact:
The global correlations between the local basis choices ($Z$ or $X$) and the local basis states ($\ket{0}$ or $\ket{+}$ \textit{vs.\/}~$\ket{1}$ or $\ket{-}$) in the SHIFT ensemble are exactly the correlations between local inputs ($a,b,c\in\{0,1\}$) and local outputs ($x,y,z\in\{0,1\}$) specified by the AF/BW process. This mathematical fact allows us to use the AF/BW process to implement the SHIFT measurement via local operations and, conversely, to use any implementation of the SHIFT measurement to realize the classical channel underlying the AF/BW process. Indeed, this observation holds more generally for multiqubit instances of QNLWE, as we now demonstrate.

\textit{Multipartite QNLWE.---}We show that {\em all\/} Boolean classical processes that violate causal order in a maximal sense---classical processes where {\em each\/} party can receive a~signal from at least one other party---give rise to ensembles that exhibit quantum nonlocality without entanglement.
Classical processes are characterized by a~{\em unique fixed-point\/} condition~\cite{baumeler2016FP,baumeler2021} as follows.
Let~$\omega^n$ be a~Boolean function~$\{0,1\}^n\rightarrow\{0,1\}^n$,
and~$\mathcal F$ the set of all functions~$\{0,1\}\rightarrow\{0,1\}$.
The function~$\omega^n$ is a~{\em Boolean~$n$-party classical process\/} if and only if
\begin{align}
	\forall \mu\in\mathcal F^n,
	\,\exists ! \underline p\in\{0,1\}^n:
	\underline p= \omega^n(\mu(\underline p))\,,
	\label{eq:ufp}
\end{align}
\textit{i.e.,} if and only if for {\em each\/} choice of interventions~$\mu_i$ of each party there exists a {\em unique\/} fixed-point of~\mbox{$\omega^n\circ\mu$}.
Here,~$\mu=(\mu_1,\mu_2,\dots,\mu_n)$ is an $n$-tuple of local Boolean functions.
Moreover, we say that~$\omega^n$ {\em has no global past\/} if and only if
\begin{align}
	\forall i\,\exists k,\underline x\in\{0,1\}^n:
	\omega^n_i(\underline x) \neq \omega^n_i(\underline x^{(k)})
	\,,
	\label{eq:maxviol}
\end{align}
where~$\underline x^{(k)}=(x_1,\dots,x_{k-1},x_k\oplus 1,x_{k+1},\dots,x_n)$ is the same as~$\underline x$ but where the~$k$-th bit is flipped, and where~$\omega^n_i$ is the~$i$-th component of~$\omega^n$.
This condition states that every party~$i$ can receive a signal through the process from at least one other party~$k$;
no party lies in the global past of all other parties.\\
\textit{Theorem.---}
If~$\omega^n$ is a Boolean~$n$-party classical process without global past,
then
\begin{align}
	\mathcal S_{\omega^n}:=
	\left\{
		H^{(\omega^n(\underline x))}\ket{\underline x} \mid \underline x\in\{0,1\}^n
	\right\}
\end{align}
is a basis of orthonormal states that exhibits QNLWE.\\

\textit{Proof.---}
The states in the set~$S_{\omega^n}$ with cardinality~$2^n$ are normalized.
Now we show that they are orthogonal, {\it i.e.,\/}
\begin{align}
	\forall \underline x\neq \underline y: \bra{\underline y}H^{(\omega^n(\underline y)\oplus\omega^n(\underline x))}\ket{\underline x} = 0\,,
	\label{eq:orthogonality}
\end{align}
where~$\oplus$ is bitwise addition modulo $2$. Pick two~\mbox{$n$-bit} strings~$\underline x\neq \underline y$ and suppose without loss of generality that they differ in the first~$k$ positions only.
Orthogonality~(Eq.~\eqref{eq:orthogonality}) states that there exists some~\mbox{$i\leq k$} with~$\omega_i^n(\underline x) = \omega_i^n(\underline y)$.
Towards a contradiction, however, assume~\mbox{$\forall i\leq k:\omega_i^n(\underline x) \neq \omega_i^n(\underline y)$}.
Since~$\omega^n$ is a classical process, the {\em reduced function\/}~$\tilde\omega^n:\{0,1\}^k\rightarrow\{0,1\}^k$ with
\begin{align}
	\underline z \mapsto (\omega^n_1(\underline z,x_{k+1},\dots, x_{n}), \dots, \omega^n_k(\underline z, x_{k+1},\dots,x_{n}))
\end{align}
is a classical process as well (see the Appendix or~\cite[Lemma~A.3]{baumeler2019}).
To simplify notation, let~$\underline x'$ be the first~$k$ bits of~$\underline x$, and similarly for~$\underline y'$, and define~\mbox{$\underline a := \tilde\omega^n(\underline x')$},~\mbox{$\underline b:= \tilde\omega^n(\underline y')$}.
Now,~$\underline a$ and~$\underline b$ are {\em fixed-points\/} under the following two~$k$-party interventions~$\alpha$ and~$\beta$, respectively, {\it i.e.,\/}~\mbox{$\underline a = \tilde\omega^n(\alpha(\underline a))$},~\mbox{$\underline b = \tilde\omega^n(\beta(\underline b))$} for
\begin{align}
	\alpha,\beta:\{0,1\}^k&\rightarrow\{0,1\}^k\in\mathcal F^k\\
	\alpha: \underline w &\mapsto \underline x'\oplus \underline a \oplus \underline w\\
	\beta: \underline w &\mapsto \underline y'\oplus \underline b \oplus \underline w
	\,.
\end{align}
However, because~$\forall i \leq k:\underline x'_i\oplus\underline y'_i=\underline a_i\oplus\underline b_i = 1$, the function~$\tilde\omega^n\circ\alpha$ has {\em a second fixed-point\/}~$\underline b$
\begin{align}
	\tilde\omega^n(\alpha(\underline b))
	&=  \tilde\omega^n(\underline x'\oplus \underline a \oplus \underline b) =  \tilde\omega^n(\underline y'\oplus \underline b \oplus \underline b)\\
	&=  \tilde\omega^n(\beta(\underline b)) = \underline b
	\,,
\end{align}
and therefore~$\omega^n$ is {\em not\/} a classical process.
This proves that the set~$\mathcal S_{\omega^n}$ forms a basis of orthonormal states.
What remains to show is that this set exhibits QNLWE.
This follows from the assumption that~$\omega^n$ has no global past.
From Eq.~\eqref{eq:maxviol} we have that for each party~$i$ there exist two bit-strings~$\underline x,\underline y$ such that the~\mbox{$i$-th} qubit of~$H^{(\omega^n(\underline x))}\ket{\underline x}$ is in the computational basis $\{\ket 0, \ket 1\}$, while the~$i$-th qubit of~$H^{(\omega^n(\underline y))}\ket{\underline y}$ is in the Hadamard basis $\{\ket +, \ket -\}$. This means that each party must change its basis depending on the bases of the other parties at least once: it follows that no party makes a basis choice that is independent of the other parties' measurements. Therefore, in an LOCC protocol for perfect discrimination, no party can initiate the communication.
\hfill$\blacksquare$

\textit{Examples.---}The following is an ensemble exhibiting QNLWE for four parties.
It is constructed from the classical process of Ref.~\cite{baumeler2022} inspired by the Ardehali-Svetlichny nonlocal game~\cite{svetlichny1987,ardehali1992}:
\begin{align}
	\begin{split}
		\big\{
		&\ket{0000}, \ket{0{+}01}, \ket{{+}01{+}}, \ket{001{-}},\\
		&\ket{01{+}0}, \ket{{+}{-}01}, \ket{01{-}0}, \ket{0111},\\
		&\ket{1{+}0{+}}, \ket{1{+}{+}{-}}, \ket{{-}01{+}}, \ket{1{+}{-}{-}},\\
		&\ket{1{-}00}, \ket{{-}{-}01}, \ket{111{+}}, \ket{1{-}1{-}}
		\big\}\,.
	\end{split}
	\label{eq:example1}
\end{align}
Another example based on the generalizations of the AF/BW process proposed in Ref.~\cite{araujo2017} is the following:
\begin{align}
	\begin{split}
		\big\{
		&\ket{0000}, \ket{0101}, \ket{0111}, \ket{1010},\\
		&\ket{1011}, \ket{1101}, \ket{1110}, \ket{1111},\\
		&\ket{001{+}}, \ket{001{-}}, \ket{01{+}0}, \ket{01{-}0},\\
		&\ket{1{+}00}, \ket{1{-}00}, \ket{+001}, \ket{{-}001}
		\big\}\,.
	\end{split}
	\label{eq:example2}
\end{align}

\textit{Conclusions.---}We have shown that Boolean \mbox{$n$-party} classical processes without global past can be mapped to a~family of $n$-qubit ensembles exhibiting quantum nonlocality without entanglement (QNLWE) and, as such, can discriminate these ensembles via local quantum operations.
We illustrated this connection explicitly for the tripartite case of the SHIFT ensemble~\cite{bennett1999} with respect to the AF/BW process~\cite{af,baumeler2016}.
This discovery therefore refines the notion of QNLWE:
Ensembles of QNLWE consist of mutually orthogonal product states that cannot be perfectly discriminated with LOCC {\em under a definite causal order.}
 
Several open questions arise from our results. 
We have, in particular, not discussed bipartite instances of quantum nonlocality without entanglement, \textit{e.g.,} the two-qutrit domino states~\cite{bennett1999}. This is because in the bipartite case, logically consistent classical processes have a definite causal order, as shown by Oreshkov \textit{et al.}~\cite{oreshkov2012}. To be sure, this instance of QNLWE {\em can\/} be interpreted as an instance of classical communication without causal order~\cite{aok17,akibue22}, but this requires a relaxation of the constraint of logical consistency which is central to the process-matrix framework \cite{oreshkov2012}. Indeed, in the bipartite case, Akibue \textit{et al.}~\cite{aok17} show that the set of transformations achievable via local operations and classical communication without causal order~(their~`LOCC*') coincides with the set of separable operations. This means that the two-qutrit domino states can be perfectly discriminated by LOCC*, as shown explicitly in~Ref.~\cite{aok17}.\footnote{Akibue in his PhD thesis \cite{akibue22} also considers a tripartite example, namely, the tripartite classical-probabilistic process proposed in Ref.~\cite{baumeler2014ieee} and shows that it can realize a non-LOCC separable operation. However, unlike the bipartite case of domino states discussed in Ref.~\cite{aok17}, this example does not admit a straightforward interpretation in terms of quantum nonlocality without entanglement.} Hence, while bipartite instances of QNLWE can be achieved under an arbitrary relaxation of causal order (as represented by LOCC*), it cannot be achieved under a relaxation of causal order that is consistent with the process-matrix framework. Our results, on the other hand, show that multipartite instances of QNLWE \textit{can} be achieved under a relaxation of causal order that is consistent with the process-matrix framework, \textit{i.e.}, without the possibility of logical paradoxes.

In the multipartite case, our results allow us to reinterpret the phenomenon of QNLWE as an operational witness of noncausality that has a qualitatively different character than the violation of causal inequalities. This opens up several potential connections with the wider literature on QNLWE and calls for a deeper understanding of its connection with noncausality. Indeed, as we have demonstrated, these results also offer a route to construct new instances of QNLWE.
These instances are of relevance for quantum cryptography, {\it e.g.,\/}~in quantum data hiding~\cite{divincenzo2002}.\footnote{Let $S$ be such an ensemble constructed out of a Boolean $n$-party classical process without global past. Given an arbitrary bit string of length $n$, we can associate it to the corresponding quantum state in $S$ and distribute the qubits to the $n$ parties. For the parties, then, it is impossible to identify the bit-string unless they meet.}
We also know that in standard quantum theory, multiqubit instances of QNLWE are incapable of witnessing a strong form of nonclassicality, \textit{i.e.,} logical proofs of the Kochen-Specker theorem \cite{wright2021contextuality}, and it would be interesting to investigate the implications of this fact for (non)causality in the process-matrix framework \cite{oreshkov2012}.
Similarly, higher-dimensional generalizations of multipartite QNLWE~\cite{niset2006} could also inspire new types of noncausal classical processes.
The domino states, however, suggest that a mapping from ensembles of QNLWE to noncausal classical processes is in general impossible.
The bipartite case, together with other generalizations of our results in the multipartite setting---in particular, the gap between separable and LOCC operations---will be taken up in forthcoming work.

The tradeoff between causal order and locality has also been studied in other senses.
Costa de Beauregard~\cite{debeauregard1977} explains entanglement through retarded waves (see also Price~\cite{price1994}),
and Deutsch's time-travel model~\cite{deutsch1991} can be turned into a local-realistic hidden-variable model for Bell correlations~\cite{baumeler2018a}.
However, this latter approach---just as the results by Akibue \textit{et al.}~\cite{aok17}---diverts from the process-matrix framework, and therefore predicts non-linear statistics, allowing the possibility of detecting new physics locally.
In contrast, processes from the process-matrix framework do not alter local physics by design, \textit{e.g.}, they do not allow signalling from the output of a party to its input.
However, if one requires the correlations to be non-signaling under any choice of interventions, then unlike QNLWE, Bell nonlocality is unaffected by any relaxation in causal order that is consistent with the process-matrix framework.

Let us also remark that whether the AF/BW process arises in general relativity would affect the interpretation of the noncausality witnessed via the perfect state discrimination task we have considered. In a Minkowski spacetime, three parties cannot discriminate the SHIFT ensemble with local operations and classical communication. However, if the parties are situated in a general-relativistic spacetime that realizes the AF/BW process, then this task becomes feasible. A successful discrimination of the SHIFT ensemble would then be an {\em operational signature\/} for the noncausal nature of such a~general-relativistic spacetime.
On the other hand, if the AF/BW process turns out not to be realizable in a general-relativistic spacetime but instead requires an~intrinsically non-classical notion of spacetime (arising from, \textit{e.g.,} quantum gravity),
then this discrimination task would serve as an operational signature of noncausality that is intrinsically non-classical.
To be sure, in such a situation, the communication between the labs would still be classical but the physical conditions for achieving this communication would be outside the realm of possibilities afforded by general-relativistic spacetimes.
The latter possibility could have interesting implications for how one might interpret time-delocalized realizations~\cite{oreshkov2019} of the AF/BW process~\cite{wechs2023existence}.

\textit{Acknowledgments.--}We thank Nicolas Cerf, Stefano Pironio, Mio Murao, Ognyan Oreshkov, and Eleftherios Tselentis for helpful discussions and comments, and two anonymous referees for their helpful comments.
RK also thanks ETH Z\"urich and IQOQI-Vienna, and \"AB also thanks QuIC for supporting the visits that made this work possible.
RK is supported by the Chargé de Recherche fellowship of the Fonds de la Recherche Scientifique FNRS~(F.R.S.-FNRS), Belgium.
\"AB~is supported by the Austrian Science Fund~(FWF) through projects ZK3 (Zukunftskolleg) and BeyondC-F7103, and by the Swiss National Science Foundation (SNF) through project~214808.
\section{Appendix: Classical processes}
The process-matrix framework~\cite{oreshkov2012} describes the most general interconnections among various parties under the assumption that it is {\em impossible\/} for the parties to locally detect any deviation from quantum theory.
Crucially, no restriction on the causal relations among the parties is made.
Classical processes, as invoked in this Letter, arise as the classical limit of the process-matrix framework~\cite{baumeler2016}, but can also be derived independently and without referring to quantum theory, as it is done in this Appendix.

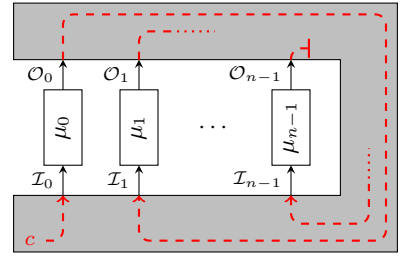
\begin{figure}
	\centering
	\begin{tikzpicture}
		\def\pw{0.5}
		\def\ph{1}
		\def\pd{.5}
		\def\offset{2/5}
		\def\Wsize{.75}
		\draw (0,0) rectangle node[rotate=90] {$\mu_0$} ++(\pw,\ph);
		\draw (\pw+\pd,0) rectangle node[rotate=90] {$\mu_1$} ++(\pw,\ph);
		\path (2*\pw+2*\pd,0) rectangle node {$\cdots$} ++(\pw,\ph);
		\draw (3*\pw+3*\pd,0) rectangle node[rotate=90] {$\mu_{n-1}$} ++(\pw,\ph);
		\draw[fill=lightgray] (-\offset,\ph+\offset) -| ++(4*\pw+3*\pd+2*\offset,-2*\offset-\ph) -| ++(-4*\pw-3*\pd-2*\offset,-\Wsize) -| ++(4*\pw+3*\pd+2*\offset+\Wsize,2*\Wsize+2*\offset+\ph) -| cycle;
		\draw[-stealth] (0*\pw+0*\pd+0.5*\pw, \ph) -- node[midway,left] {\scriptsize$\mathcal O_0$} ++(0,\offset);
		\draw[-stealth] (1*\pw+1*\pd+0.5*\pw, \ph) -- node[midway,left] {\scriptsize$\mathcal O_1$} ++(0,\offset);
		\draw[-stealth] (3*\pw+3*\pd+0.5*\pw, \ph) -- node[midway,left] {\scriptsize$\mathcal O_{n-1}$} ++(0,\offset);
		\draw[stealth-] (0*\pw+0*\pd+0.5*\pw, 0) -- node[midway,left] {\scriptsize$\mathcal I_0$} ++(0,-\offset);
		\draw[stealth-] (1*\pw+1*\pd+0.5*\pw, 0) -- node[midway,left] {\scriptsize$\mathcal I_1$} ++(0,-\offset);
		\draw[stealth-] (3*\pw+3*\pd+0.5*\pw, 0) -- node[midway,left] {\scriptsize$\mathcal I_{n-1}$} ++(0,-\offset);
		\def\bstep{4/5}
		\def\lstep{1/2}
		\def\mstep{1/5}
		\draw[red,thick,dashed,rounded corners,<-] (.5*\pw,-\offset) |- ++(-.5*\pw,-\bstep*\Wsize) node[left] {\footnotesize$c$};
		\draw[red,thick,dashed,rounded corners,->] (.5*\pw,\ph+\offset) |- ++(3*\pd+3.5*\pw+\offset+\bstep*\Wsize,\bstep*\Wsize) |- ++(-2*\pd-2.5*\pw-\offset-\bstep*\Wsize,-2*\bstep*\Wsize-\ph-2*\offset) -- ++(0,\bstep*\Wsize);
		\draw[red,thick,dashed,rounded corners,-] (\pw+\pd+.5*\pw,\ph+\offset) |- ++(\pw,\lstep*\Wsize);
		\draw[red,thick,dotted,rounded corners,-] ($ (\pw+\pd+.5*\pw,\ph+\offset) + (\pw,\lstep*\Wsize) $) -- ++(\pw,0);
		\draw[red,thick,dashed,rounded corners,<-] (3*\pw+3*\pd+.5*\pw,-\offset) |- ++(.5*\pw+\offset+\lstep*\Wsize,-\lstep*\Wsize) -- ++(0,\pw);
		\draw[red,thick,dotted,rounded corners,-] ($ (3*\pw+3*\pd+.5*\pw,-\offset) + (.5*\pw+\offset+\lstep*\Wsize,-\lstep*\Wsize) + (0,\pw) $) -- ++(0,\pw);
		\draw[red,thick,dashed,rounded corners,-|] (3*\pw+3*\pd+.5*\pw,\ph+\offset) |- ++(+.5*\pw,\mstep*\Wsize);
	\end{tikzpicture}
	\caption{Each party~$k\in[n]$ implements a~function~$\mu_k:\mathcal I_k\rightarrow\mathcal O_k$ of their choice.
		The process, {\it i.e.,}~the grayed-out higher-order map, interconnects the parties.
	In red, we have schematically displayed an example where party~$0$ receives a~constant~$c$, party~$1\leq k\leq n-1$ receives the output of party~$k-1$, and the output of party~$n-1$ is discarded:
	A~corresponding process is~$\omega:\underline{\mathcal O}\rightarrow\underline{\mathcal I}$ with~$\omega(o_0,o_1,\dots,o_{n-1})=(c,o_0,\dots,o_{n-2})$.
}
	\label{fig:description}
\end{figure}

For the following description of the classical processes, you may consult Fig.~\ref{fig:description}.
Consider an~$n$-party scenario, where we label the parties with the natural numbers~\mbox{$[n]:=\{0,1,\dots,n-1\}$}.
Each party~$k\in[n]$ is formalized as a pair of sets~$(\mathcal I_k,\mathcal O_k)$.
The set~$\mathcal I_k$ is the {\em input space\/} of~$k$, and~$\mathcal O_k$ the {\em output space\/} of party~$k$.
Moreover, we define~$\mathcal F_k:=\{\mathcal I_k\rightarrow\mathcal O_k\}$ as the set of all functions from the input space to the output space of party~$k$.
The assumptions of the framework are: (A)~Each party~$k\in[n]$ can implement any function (hereafter called {\em intervention})~$\mathcal \mu_k\in\mathcal F_k$ of their choice, (B)~the parties are isolated and may only communicate by reading messages from the input spaces and inscribing messages to the output spaces, and (C) each party~$k\in[n]$ gets an input~$i_k\in\mathcal I_k$ {\em exactly once\/} and applies the chosen intervention~$\mu_k$ {\em exactly once.}
For the sake of presentation, we define the Cartesian product~$\underline{\mathcal I}:=\bigtimes_{k\in[n]}\mathcal I_k$, and similarly for~$\underline{\mathcal O}$ and~$\underline{\mathcal F}$.
Also, we define the collection~$\underline{i}:=(i_k)_{k\in[n]}$ and similarly for~$\underline{o}$.
Assumptions~(B) and~(C) require~$\underline{i}$ to functionally depend on~$\underline{o}$, {\it i.e.,}~$\omega(\underline{o})=\underline{i}$, for some function~$\omega:\underline{\mathcal O}\rightarrow\underline{\mathcal I}$.
The value of~$\underline{o}$, then again, functionally depends on~$\underline{i}$ through the choice of interventions~$\mu$, {\it i.e.,}~$\underline{o}=\mu(\underline{i})$.
By invoking assumption (A), an {\em $n$-party process\/} is a~function~$\omega:\underline{\mathcal O}\rightarrow\underline{\mathcal I}$ that satisfies the fixed-point condition
\begin{align}
	\forall \mu\in\underline{\mathcal F},\;\exists \underline{i}\in\underline{\mathcal I}: \underline{i}=\omega(\mu(\underline{i}))
	\,.
	\label{eq:fp}
\end{align}
Thus, for any choice of interventions~$\mu$ of the parties, a~well-defined input~$\underline{i}$ to the parties exists.
This fixed-point condition~\eqref{eq:fp} has as consequence~\cite{baumeler2021} the {\em unique fixed-point condition\/} (Eq.~\eqref{eq:ufp} in this Letter):
\begin{align}
	\forall \mu\in\underline{\mathcal F},\;\exists! \underline{i}\in\underline{\mathcal I}: \underline{i}=\omega(\mu(\underline{i}))
	\,,
\end{align}
where~$\exists!$ is the {\em uniqueness\/} quantifier.
This, actually, ensures {\em logical consistency:}
The input~$\underline{i}$ to the parties is unambiguously determined (see Ref.~\cite{baumeler2021} for a detailed discussion).

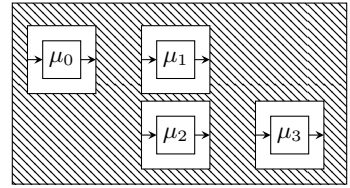
\begin{figure}
	\centering
	\begin{tikzpicture}
		\def\ps{.5}
		\def\offset{.4}
		\draw[pattern=north west lines] (0,0) rectangle ++(4+\offset,2+\offset);
		\def\coordx{0} \def\coordy{1} \def\party{$\mu_0$}
		\draw[fill=white] (\offset+\coordx-.5*\offset,\offset+\coordy-.5*\offset) rectangle ++(\ps+\offset,\ps+\offset);
		\draw             (\offset+\coordx,\offset+\coordy) rectangle node {\party} ++(\ps,\ps);
		\draw[-stealth]   (\offset+\coordx+\ps,\offset+\coordy+.5*\ps) -- ++(.5*\offset,0);
		\draw[stealth-]   (\offset+\coordx,\offset+\coordy+.5*\ps) -- ++(-.5*\offset,0);

		\def\coordx{1.5} \def\coordy{1} \def\party{$\mu_1$}
		\draw[fill=white] (\offset+\coordx-.5*\offset,\offset+\coordy-.5*\offset) rectangle ++(\ps+\offset,\ps+\offset);
		\draw             (\offset+\coordx,\offset+\coordy) rectangle node {\party} ++(\ps,\ps);
		\draw[-stealth]   (\offset+\coordx+\ps,\offset+\coordy+.5*\ps) -- ++(.5*\offset,0);
		\draw[stealth-]   (\offset+\coordx,\offset+\coordy+.5*\ps) -- ++(-.5*\offset,0);

		\def\coordx{1.5} \def\coordy{0} \def\party{$\mu_2$}
		\draw[fill=white] (\offset+\coordx-.5*\offset,\offset+\coordy-.5*\offset) rectangle ++(\ps+\offset,\ps+\offset);
		\draw             (\offset+\coordx,\offset+\coordy) rectangle node {\party} ++(\ps,\ps);
		\draw[-stealth]   (\offset+\coordx+\ps,\offset+\coordy+.5*\ps) -- ++(.5*\offset,0);
		\draw[stealth-]   (\offset+\coordx,\offset+\coordy+.5*\ps) -- ++(-.5*\offset,0);

		\def\coordx{3} \def\coordy{0} \def\party{$\mu_3$}
		\draw[fill=white] (\offset+\coordx-.5*\offset,\offset+\coordy-.5*\offset) rectangle ++(\ps+\offset,\ps+\offset);
		\draw             (\offset+\coordx,\offset+\coordy) rectangle node {\party} ++(\ps,\ps);
		\draw[-stealth]   (\offset+\coordx+\ps,\offset+\coordy+.5*\ps) -- ++(.5*\offset,0);
		\draw[stealth-]   (\offset+\coordx,\offset+\coordy+.5*\ps) -- ++(-.5*\offset,0);
	\end{tikzpicture}
	\caption{A circuit with holes is a process.}
	\label{fig:circuit}
\end{figure}

Processes are generalizations of shared states, communication channels, and circuits.
A shared state---where the parties do not communicate---is simply given by a~process
\begin{align}
	\omega_\text{state}:\underline{o}\mapsto (c_0,c_1,\dots,c_{n-1})
	\,,
\end{align}
for some constants~$(c_k)_{k\in[n]}$.
Here, the fixed-point condition is satisfied independently from the choice of interventions~$\mu$ by the constants~$(c_k)_{k\in[n]}$.
An example of a communication channel, as schematically depicted in Fig.~\ref{fig:description}, is given by
\begin{align}
	\omega_\text{com}:(o_0,o_1,\dots,o_{n-1})\mapsto (c,o_0,o_1,\dots,o_{n-2})
	\,,
\end{align}
for some constant~$c$.
This communication channel provides the constant~$c$ to party~$0$, and each remaining party~$k$ obtains~$o_{k-1}$ on their input space.
Here, the fixed-point {\em depends\/} on the choice of interventions.
It is given by~$i_0 = c$ for party~$0$, and~$i_k = \mu_{k-1}\circ\cdots\circ\mu_1\circ\mu_0(c)$ for each remaining party~$k$.
More complex situations are also expressible with processes.
For instance, take a circuit~$\mathcal C$ composed out of classical gates, and now let each party~$k\in[n]$ occupy the region of a gate in~$\mathcal C$ (see Fig.~\ref{fig:circuit}).
Here, the process~$\omega_{\mathcal C}$ simply implements the transformations on the non-occupied regions of~$\mathcal C$.

\subsection{Classical communication without definite causal order}
Classical processes for three or more parties allow for scenarios {\em beyond\/} those discussed above.
In the above examples, a {\em global\/} causal ordering of the parties always exists.
This is radically contrasted with the AF/BW process~\cite{af,baumeler2016} (Eq.~\eqref{eq:afbw} in this Letter):
\begin{align}
	\omega_{\text{AF/BW}}(x,y,z) = \left( 
		(y\oplus 1)z,
		(z\oplus 1)x,
		(x\oplus 1)y
	\right)
	\,.
	\label{eq:Wafbw}
\end{align}
To see this, we can devise causal inequalities---similar to Bell inequalities~\cite{bell1964}---that limit the possible correlations among the parties under the assumption of a global causal order.
Let~$P(a,b,c|x,y,z)$ be three-party correlations where a party---say Alice---specifies a setting~$x$ and observes the outcome~$a$, and similarly for the other two parties Bob and Charlie.
The assumption of a global causal order limits the parties to only influence events in their causal future.
So, three-party correlations are called {\em causal\/} if and only if they can be decomposed as
\begin{align}
	P(a,b,c|x,y,z) &= 
	\lambda_A P(a|x) P_{a,x}(b,c|y,z)\\
	+
	&\lambda_B P(b|y) P_{b,y}(a,c|x,z)\\
	+
	&\lambda_C P(c|z) P_{c,z}(a,b|x,y)
	\,,
\end{align}
with~$\lambda_A,\lambda_B,\lambda_C\geq 0$,~$\lambda_A+\lambda_B+\lambda_C=1$, and where~$P_{a,x}(b,c|y,z)$ and the other terms denote two-party causal correlations.
Here,~$\lambda_A$ specifies the probability that Alice acts first.
Recursively, two-party correlations~$P(a,b|x,y)$ are {\em causal\/} if and only if they can be decomposed as
\begin{align}
	P(a,b|x,y)
	&=
	\gamma
	P(a|x)P(b|y,a,x)\\
	+
	&(1-\gamma)
	P(b|y)P(a|x,b,y)
	\,,
\end{align}
for some~$\gamma\geq 0$.

Let~$a,b,c,x,y,z$ be the values of binary random variables.
If the three-party correlations~$P(a,b,c|x,y,z)$ are {\em causal,} then they satisfy the following causal inequality for uniformly distributed~$x,y,z$~\cite{baumeler2016}:
\begin{align}
	\Pr[(a,b,c) = \omega_{\text{AF/BW}}(x,y,z)] \leq 3/4
	\,.
\end{align}
Clearly, this inequality is deterministically violated whenever Alice, Bob, and Charlie communicate through the AF/BW process:
The AF/BW process allows for correlations incompatible with any global causal order of the parties.

\subsection{Reduced processes}
In this Letter we make use of {\em reduced functions.}
Consider a function~\mbox{$\omega:\bigtimes_{k\in[n]}\mathcal O_k \rightarrow\bigtimes_{k\in[n]}\mathcal I_k$} with its components~$\{\omega_\ell:\bigtimes_{k\in[n]}\mathcal O_{k} \rightarrow\mathcal I_\ell\}_{\ell\in[n]}$.
We call the function~$\omega$ {\em component-wise non-signalling\/} if and only if for all~$\ell$, the~$\ell$-th input to~$\omega$ does not influence the output of the component~$\omega_\ell$, {\it i.e.,}
\begin{align}
	\forall \ell\in[n],\; &(o_\ell,o'_\ell)\in\mathcal O_\ell^2,\;o_{\setminus\ell}\in\!\bigtimes_{k\in[n]\setminus\{\ell\}}\!\mathcal O_k:
	\\
	\omega_\ell&(o_\ell,o_{\setminus\ell})
	=
	\omega_\ell(o'_\ell,o_{\setminus\ell})
	\,.
\end{align}
If the function~$\omega$ is component-wise non-signaling, we can define the {\em reduced function\/}~$\omega^{\mu_r}$ for all parties~$\ell\neq r$, where the intervention~$\mu_r$ of one party~$r$ is taken into account
\begin{align}
	\omega_\ell^{\mu_r}:\bigtimes_{k\in[n]\setminus\{r\}}\mathcal O_k &\rightarrow \mathcal I_\ell\\
	(o_k)_{k\in[n]\setminus\{r\}} &\mapsto \omega_\ell(\dots,o_{r-1},\hat o_r,o_{r+1},\dots)
\end{align}
by specifying
\begin{align}
	\hat o_r := \mu_r\circ\omega_r(\dots,o_{r-1},c_r,o_{r+1},\dots)
\end{align}
for some {\em arbitrary\/}~$c_r$.
This allows for the following statement~\cite{baumeler2019}.
If~$\omega$ is an~$n$-party process, then~$\omega$ is {\em component-wise non-signaling,} and for all parties~$r\in[n]$ and all interventions~$\mu_r\in\mathcal F_r$, the reduced function~$\omega^{\mu_r}$ is an~$(n-1)$-party process.

\bibliography{refs.bib}
\end{document}